\documentclass[12pt,a4paper,english]{article}
\usepackage{graphicx}
\usepackage{epsf}
\usepackage{amsmath}
\usepackage{amssymb}
\usepackage[T2A]{fontenc}
\usepackage[cp1251]{inputenc}
\usepackage{babel}
\usepackage[labelsep=period,hang]{caption} 
\newcommand{\NMS}[1]{\mbox{$#1\,M_{\textstyle\odot}$}}
\newcommand{\ZMS}[1]{\mbox{$#1\,Z_{\textstyle\odot}$}}
\newcommand{\xmn}[2]{\mbox{$#1\!\times\! 10^{#2}\,$}}
\newcommand{\isn}[2]{\mbox{$^{#2}${#1}}}
\newcommand{\avr}[1]{\mbox{$\langle #1\rangle$}}
\newcommand{\indis}[1]{{\mbox{\scriptsize #1}}}

\newcommand{\nucmu}{\mbox{$m_{\mathrm u}$}}

\newcommand{\fradd}[2]{\mbox{$\frac{\textstyle\mathrm{d}#1}{\textstyle\mathrm{d}#2}$}}

\title{Neutrino-induced nucleosynthesis as a result of
mixing between the He and C-O-Ne shells
in core-collapse supernovae}

\author{D.K.~Nadyozhin$^{1,2}$ and I.V.~Panov$^{1,2,3,4}$%
        \thanks{E-mail: {nadezhin@itep.ru}, {Igor.Panov@itep.ru}}\\
{\it  $^{1}$Institute for Theoretical and Experimental Physics, Moscow, Russia}\\
{\it  $^{2}$SRC Kurchatov Institute, Moscow, Russia}\\
{\it  $^{3}$Novosibirsk State University, Novosibirsk, Russia} \\
{\it  $^{4}$Moscow Institute of Physics and Technology, Dolgoprudny, Russia}}

\begin{document}

\date{}

\maketitle
\abstract{
 The nucleosynthesis yields from neutrino-induced spallation
 of \isn{He}{4} are calculated in case of penetration of
 some \isn{He}{4} into the carbon-oxygen-neon shell.
 The neutrino contribution to creation of the weak r-process
 component becomes noticeable in case \isn{He}{4} is dragged
 down at radii $R\lesssim 10^9\,$cm.}

\section*{Introduction}
 The interaction of neutrino flux with \isn{He}{4}
 in supernova He-shell was first discussed in \cite{DomEramNad78}
 and used there to estimate the synthesis of light elements such as Li, Be, and B.
 Then the neutrino-induced disintegration of \isn{He}{4} was suggested
 to be a source of neutrons for driving the r-process \cite{EpsteinColgateHaxton88}.

 Following studies of this suggestion revealed that
 the resulting neutrino-induced yields of heavy nuclides were sensitive
 to a number of parameters describing the physical condition in presupernova
 He-shell and depend on not yet finally established details of
 the supernova mechanism (see \cite{NPB98}--\cite{NadPan2008} and references therein).

 The most important parameters are:
  (i) the neutrino ``light curve'' and the neutrino energy and flavor
       spectra;
  (ii) the supernova total explosion energy $E_{\mathrm{exp}}$ that
       determines the strength of the shock wave and time it takes
       to propagate through the presupernova envelope;
  (iii) The properties of the onion-like presupernova chemical structure,
       especially the radii $R$ of different chemical shells.

  There was shown \cite{NadPan2007} that neutrons produced by neutrino interaction
  with \isn{He}{4} could appreciably contribute to a light (weak) component of
  a two-component r-process model \cite{WBG1996,QW2000}, especially
  in case of a low metallicity presupernova.
  However, this becomes possible if either the number densities
  of the neutron poisons in He-shell,
  such as \isn{C}{12}, \isn{N}{14}, \isn{O}{16}, \isn{Ne}{20}, do not
  at least exceed that of iron seeds or the radius of He-shell is below \xmn{(1-2)}{9}cm.
  Both the constraints are in conflict with currently available presupernova models
  in which some neutron poison abundances exceed by number that of iron seeds
  by a factor of 10--100 and the He-shell radii are in the interval
  \xmn{4}{9}-- \xmn{4}{10}cm.

  The calculations in \cite{NadPan2007} demonstrated the possibility of
  the neutrino-induced creation of the weak r-process component $(A\lesssim 130)$
  for the He-shell radius \xmn{1.37}{10}cm when admixtures of neutron poisons
  are deactivated. These admixtures can depend on the still poorly
  studied diffusion and mixing of matter at the star’s final evolutionary phases
  (semiconvection, meridional circulation).

  The present work deals with the possibility to obtain
  the weak r-process component by means of decreasing the He-shell radius.
  We  suggest that shortly before the beginning of gravitational collapse
  large scale circulation of matter can occur between the carbon-oxygen-neon
  and helium shells. This way a noticeable amount of helium could be transported
  down to be exposed to much stronger neutrino flux at smaller radii.
  Such suggestion assumes the braking of spherical symmetry at final stages
  of stellar evolution  that was already repeatedly discussed in literature,
  see for example \cite{BazanArnett98} and references therein.

 The preliminary results of this work were reported
 at the 16th Workshop on ``Nuclear Astrophysics'', Ringberg Castle
 (2012) \cite{Ringberg12}.

\section*{Presupernova model}

\begin{figure}[htb!]
  \centerline{\includegraphics[clip,width=0.8\textwidth]{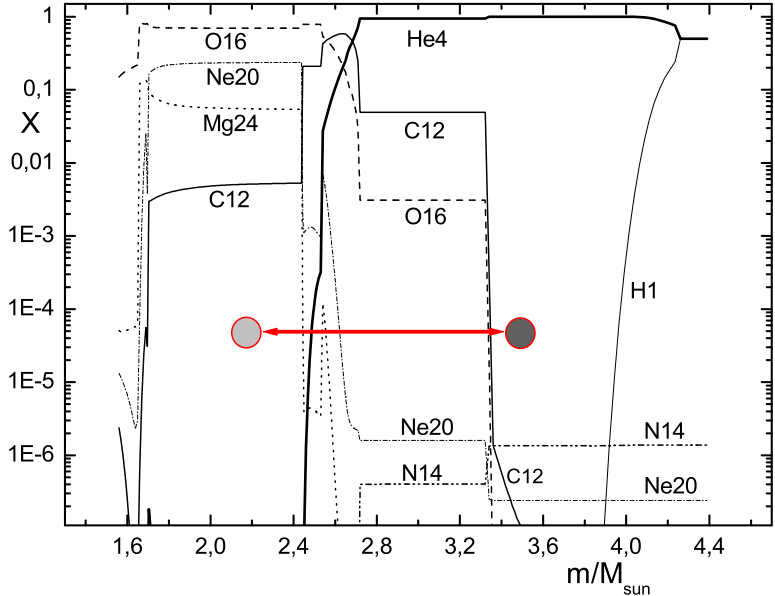}}
  \caption{The composition of the He and C-O-Ne shells
   of a \protect\NMS{15} low metallicity (\ZMS{0.0001}) presupernova model
   \cite{WoosleyHegerWeaver2002}
   }
  \label{ddz-4-15}
\end{figure}

To estimate the efficiency of mixing between the He and
C-O-Ne shells for production of the weak r-process component
we make use of a \NMS{15} low metallicity (\ZMS{0.0001})
evolutionary presupernova model.
 Labeled as dd15z-4, the model belongs to subsidiary materials used for
 preparation of the review paper \cite{WoosleyHegerWeaver2002}.
Figure$\,$\ref{ddz-4-15} shows the mass fractions $X$ of chemical elements
in the He and C-O-Ne shells versus mass coordinate $m$. Dark gray circle
in the He shell and light gray one in the C-O-Ne shell enclose the region
of mixing in our calculations. The properties of stellar matter for both
the circles are listed in Table$\,$\ref{Tab1}.
\begin{table}
  \centering
  \caption{Properties of layers undergoing mixing in presupernova
   model \mbox{dd15z-4}}\label{Tab1}
\vspace*{3mm}
  \begin{tabular}{cc}
    \hline\hline\\[-3mm]
    Mid of He shell (zone 365) & C-O-Ne shell (zone 230)\\ \hline
    & \\[-3mm]
    $m=\NMS{3.50}$             & $m=\NMS{2.15}$ \\
    $R_0=\xmn{1.58}{10}$cm     & $R_0=\xmn{1.3}{9}$cm \\
    $T=\xmn{7.914}{7}$K        & $T=\xmn{7.7}{8}$K \\
    $\rho=36.6\,$g$\,$cm$^{-3}$& $\rho=\xmn{3.9}{4}\,$g$\,$cm$^{-3}$\\
    $X_\indis{He}\approx 1.0000$     & $X_\indis{He}= 0.0 $ \\
    $X_\indis{C12}=\xmn{9.743}{-8}$  & $X_\indis{C12}=\xmn{5.089}{-3}$ \\
    $X_\indis{N14}=\xmn{1.331}{-6}$  & $X_\indis{N14}=0.0$\\
    $X_\indis{O16}=\xmn{3.201}{-8}$  & $X_\indis{O16}=\xmn{6.978}{-1}$ \\
    $X_\indis{Ne20}=\xmn{2.40}{-7}$  & $X_\indis{Ne20}=\xmn{2.353}{-1}$ \\
    $X_\indis{Mg24}=\xmn{4.11}{-8}$  & $X_\indis{Mg24}=\xmn{5.539}{-2}$ \\
    $X_\indis{Fe}=\xmn{1.38}{-7}$    & $X_\indis{Fe}=\xmn{1.38}{-7}$\\
    \hline \hline
  \end{tabular}
\end{table}

 We assume that the mixing between He and C-O-Ne shells occurs
 at the last stage of hydrostatic presupernova evolution before
 onset of a powerful neutrino burst. As a result, a portion of He from
 the He shell (dark grey circle in Fig.$\,$\ref{ddz-4-15}) proved to be
 dragged into C-O-Ne shell (light grey circle) where it undergoes mixing
 with the C-O-Ne composition. Therefore, the mixed composition reads
\begin{equation}\label{almix}
X_i =\alpha X_i(\mbox{He}) + (1-\alpha) X_i(\mbox{CONe})\, ,
\quad (i=\isn{He}{4},\isn{C}{12},\ldots,\isn{Mg}{24},\isn{Fe}{}),
\end{equation}
where $X_i(\mbox{He})$ and $X_i(\mbox{CONe})$ are weight abundances
from the left and right columns of Table$\,$\ref{Tab1}. Efficiency of mixing
is controlled in our calculations by a parameter $\alpha$. When $\alpha =0$
there is no intrusion of He shell matter at all.
In case $\alpha =1$ we have an unsullied matter
from He shell inside C-O-Ne one.
Helium should penetrate into the \mbox{C-O-Ne} shell not earlier than several
seconds before the neutrino burst. Otherwise He would be burnt in
thermonuclear reactions
(in our case, mostly in the \isn{O}{16}$(\alpha,\gamma)$\isn{Ne}{20} one)
rather than to interact with the neutrino flux.

The weight abundance of the iron group seeds for r-process in the model
dd15z-4 turns out to be as small as
$X_\indis{Fe}=\xmn{1.38}{-7}$,
$(Y_\indis{Fe}\approx X_\indis{Fe}/56 = \xmn{2.46}{-9})$.

\section*{The neutrino flux properties and neutrino-nucleus interactions}
The rate of the neutrino-induced transformation of a nuclide of
mass number $A$ is given by
\begin{equation}\label{dYdt}
\left(\fradd{Y_A}{t}\right)_\nu = -\, q_\nu\,\frac{L_{\indis{tot}}(t)}{\avr{E_\nu}}
\,\frac{\avr{\sigma_{A\nu}}}{4\pi R^2(t)}\, Y_A\, ,
\; (\nu =\nu_{\indis{e}}, \bar\nu_{\indis{e}},\nu_\mu, \bar\nu_\mu,
\nu_\tau, \bar\nu_\tau),
\end{equation}
where $Y_A= n_A\,\nucmu/\rho$ is the number of nuclides $n_A$ in unit volume
per barion, $L_{\indis{tot}}(t)$ is the total neutrino and antineutrino
luminosity of all the neutrino flavors, $q_\nu$ is the fraction of
the total neutrino flux responsible for the interaction in question,
\avr{E_\nu} is the neutrino mean individual energy, $R(t)$ is the radius
of the shell exposed to the neutrino interaction, and \avr{\sigma_{A\nu}}
is the cross-section averaged over the neutrino energy spectrum.

Integrating Eq.$\,$\ref{dYdt} by time at constant $R(t)=R_0$,
one can estimate that the magnitude of the neutrino-induced
transformations is given by
\begin{equation}\label{numag}
    \xi\equiv\left|\frac{\delta n_A}{n_A}\right|=
    N_\nu\frac{\avr{\sigma_{A\nu}}}{4\pi R_0^2}\approx(10^{-3}-10^{-5})\, ,
\end{equation}
where $N_\nu$ is total number of neutrino radiated whereas the numerical values
are for typical case $N_\nu \approx10^{58}$, \avr{\sigma_{A\nu}}
$\approx 10^{-42}$cm$^2$, $R_0= (10^9-10^{10})\,$cm. Therefore, as a rule
one can neglect the neutrino interactions with the nuclides already produced
by neutrinos. The neutrino-induced yields of such secondary interactions
are of the order of $\xi^2$.

 We use the temporal behavior of $L_{\indis{tot}}$ based on
 the calculations of gravitation collapse of a \NMS{1.82}
 iron core surrounded with a \NMS{0.18} oxygen envelope
 \cite{Nad77a}--\cite{Nad78} (see also Figure$\,$1 in \cite{NPB98}).
 Full energy carried away by neutrinos
 $\int^\infty_0 L_{\indis{tot}}(t)dt$ turned out to be \xmn{5.3}{53}erg.
 Henceforth we assume that the neutrino flux starts at $t=0$.

 In our calculations we also assume that $L_{\indis{tot}}(t)$ is
 equidistributed among all the neutrino species, i.e.
 $q_{\nu\mathrm{e}}=q_{\bar\nu\mathrm{e}}=q_{\nu\mu}=q_{\bar\nu\mu}=
 q_{\nu\tau}=q_{\bar\nu\tau}= 1/6$.
 During the first $\sim 100\,$ms of the collapse,
 $L_{\indis{tot}}(t)$ is building up mostly by the electron neutrino
 from neutronization of stellar matter. However, later on
 the approximate equidistribution over the neutrino flavors set in.

 The neutrino spectra are assumed to be the Fermi--Dirac distributions
 with zero chemical potentials and temperatures 7.94$\,$MeV for $\mu$
 and $\tau$ neutrino and antineutrino and 3.81$\,$MeV for electron
 antineutrino. The corresponding neutrino mean energies \avr{E_\nu}
 are 25$\,$MeV and 12$\,$MeV, respectively.

 We took into account the neutrino-nuclear neutral current interactions listed in
 Table$\,$\ref{Tab2} where $\nu$ stands for $\mu$ and $\tau$ neutrino
 and antineutrino.  The charged current electron antineutrino $\bar{\nu}_\mathrm{e}$
 capture by free protons of the order of $\xi^2$ was included as well.
 We neglected the
 reaction $\isn{He}{4}(\bar{\nu}_\mathrm{e},\mathrm{e}^+ n)\isn{H}{3}$
 \cite{BanHaxQuian11} since its cross-section is
 about 2 orders of magnitude less than that for neutral currents.
 The neutrino cross-sections averaged over the Fermi--Dirac
 spectra were taken from \cite{Heg2005,Woos90,DomSIm82}.

\begin{table}[htb]
\centering
 \caption{The neutrino-nuclear interactions}\label{Tab2}
\vspace{0.2cm}
\begin{tabular}{lc}
\hline\hline\\[-0.32cm]
 Reaction & Mean cross-section \\
          & \avr{\sigma_{A\nu}}$\,(10^{-42}\,$cm$^2)$\\ \hline
          & \\[-0.32cm]
\isn{He}{4}($\nu,\nu^\prime$ n)\isn{He}{3} & 0.403\\
\isn{He}{4}($\nu,\nu^\prime$ p)\isn{H}{3}  & 0.441\\
\isn{C}{12}($\nu,\nu^\prime$ n)\isn{C}{11} & 0.512\\
\isn{C}{12}($\nu,\nu^\prime$ p)\isn{B}{11} & 1.86\\
\isn{C}{12}($\nu,\nu^\prime$ \isn{He}{3})\isn{Be}{9} & 0.0024\\  
\isn{C}{13}($\nu,\nu^\prime\,\alpha$)\isn{Be}{9} & 0.714\\ 
\isn{N}{14}($\nu,\nu^\prime\,\alpha$)\isn{B}{10} & 0.312\\
\isn{O}{16}($\nu,\nu^\prime$ n)\isn{O}{15} & 0.747\\
\isn{O}{16}($\nu,\nu^\prime$ p)\isn{N}{15} & 2.68\\
\isn{Ne}{20}($\nu,\nu^\prime\,$n)\isn{Ne}{19} & 1.05\\
\isn{Ne}{20}($\nu,\nu^\prime\,$p)\isn{F}{19} & 7.31\\
\quad\, p($\bar{\nu}_\mathrm{e},\mathrm{e}^+$)n & 9.70\\[-0.4cm]
                                                 & \\ \hline\hline
\end{tabular}
\end{table}

\section*{Shock wave}
 The main parameters determining
 the temporal behavior of shocked matter are the total energy of
 the explosion $E$ which comes from the supernova mechanism,
 the initial (pre-shock) density $\rho_0$ and the radius $R_0$
 of the shell under consideration (the Si-S, C-O-Ne, or He one).
 The initial temperature $T_0$ of the shell is virtually unimportant
 since the SW is always very strong.

  The study of the shock wave (SW) propagation through the
  presupernova structure
  allowed to construct simple formulae that approximate the
  temporal behavior of the shocked matter properties resulting from
  detailed hydrodynamical calculations \cite{ND02}.
  For the C-O-Ne shell the formulae read as follows
\begin{eqnarray}
 T(t) & = & {T_\mathrm{p}\over 1 +\xi_T\, (t-t_\indis{SW})/t_\mathrm{u}}\, ,
 \quad T_\mathrm{p}\, =\,\xi_\mathrm{p}\,T_{\mbox{\scriptsize WW}}\, ,
 \label{Tt}\\
 \rho(t) & = & \rho_\mathrm{p}\left({T\over T_\mathrm{p}}\right)^3 ,
 \qquad\rho_\mathrm{p}\, =\, 7\rho_0\, ,  \label{Rhot}\\
 R(t) & = & R_0\left[1 + \xi_r (t-t_\indis{SW})/t_\mathrm{u}\right] ,\label{Rt}
 \end{eqnarray}
 where $T_\mathrm{p}$ and $\rho_\mathrm{p}$ are
 the peak temperature and density of shocked matter while
 $\xi_\mathrm{p}$, $\xi_T$, and $\xi_r$ are the dimensionless
 structural coefficients chosen to fit
 the hydrodynamic calculations as close as possible.

 Equations$\,$(\ref{Tt}--\ref{Rt}) imply that the SW
 takes time $t=t_\indis{SW}$ to reach  the layer after the
 beginning of the neutrino flux. Therefore, Eqs.$\,$(\ref{Tt}--\ref{Rt})
 are only valid for $t\geqslant t_\indis{SW}$ whereas within time
 interval $0\leqslant t< t_\indis{SW}$ the temperature, density and radius
 remain equal their initial values $T_0$, $\rho_0$, and $R_0$.

 Parameters $t_\mathrm{SW}$, $T_{\mbox{\scriptsize WW}}$
 (a Weaver--Woosley \cite{WW80} estimate of the peak temperature),
 and characteristic time $t_\mathrm{u}$ are given by
  \begin{eqnarray}
 t_\mathrm{SW}\! & = &  \delta t\, +\, \xi_\mathrm{SW}\, E^{-0.38}_{51}
 \left(R_{09}\right)^{1.4}\mbox{sec}\, ,\label{Dt}\\
 T_{\mbox{\scriptsize WW}}\! & = & \!\!\left({3E\over
 4\pi a R_0^3}\right)^{1/4}\!\! =\mbox{$2.37\!\times\! 10^{9}$}
  \, E_{51}^{0.25}\!
 \left(R_{09}\right)^{-0.75}\!\mbox{K},\label{TWW}\\
 t_\mathrm{u}\! & = &\!\mbox{$3.83\!\times\! 10^{-3}\,$}
 \rho^{0.5}_0 E_{51}^{-0.5}
  \left(R_{09}\right)^{2.5}\,\mbox{sec}\, ,
 \label{tu}
 \end{eqnarray}
 where
 $E_{51}\equiv E/10^{51}\,$erg, $R_{09}\equiv R_0/10^9\,\mbox{cm}$,
 and $\xi_\mathrm{SW}$ is another dimensionless structural coefficient.
 All such coefficients listed above slightly depend on the composition
 of the shell under consideration and presupernova mass.
 They are tabulated in \cite{ND02}.

 For the
 C-O-Ne shell of a \NMS{15} presupernova we use here
 $\xi_\mathrm{p}=0.95$, $\xi_T=1.4$, $\xi_r=1.25$, $\xi_\mathrm{SW}=1.1$.
 The term $\delta t$ in the
 right hand side of Eq.$\,$(\ref{Dt}) is the delay time necessary for
 converting of standing accretion shock into outgoing blast wave.
 A standard value $\delta t =100\,$ms was assumed in our calculations.
 In the calculations we used in Eqs.$\,$(\ref{Tt}--\ref{tu})
 standard value for the explosion
 energy $E_{51}=1$ and the values of $R_0$ and $\rho_0$ from Table$\,$\ref{Tab1}
 for the C-O-Ne shell. Finally we have
 \begin{equation}\label{swpeak}
T_\mathrm{p}=\xmn{1.85}{9}\mathrm{K},\;
\rho_\mathrm{p}=\xmn{2.75}{5}\mathrm{g}\,\mathrm{cm}^{-3}\! , \;
t_\mathrm{SW}=1.688\,\mathrm{s}, \;
t_\mathrm{u}=1.457\mathrm{s} .
 \end{equation}
 As a result, the radius $R(t)$ of the Lagrangian layer
 increases by a factor of 2 in 1.17$\,$s after the SW arrival whereas
 the temperature $T(t)$ becomes two times less than $T_\mathrm{p}$ in 0.83$\,$s.

\section*{Method of calculation}
 We have used two nuclear kinetics codes.
 The first one controls the nuclear kinetics for the light nuclides
 from D, \isn{H}{3}, \isn{He}{3}, $\ldots$ through \isn{Mg}{24} (the L code).
 The L code calculate the kinetics of about 160 thermonuclear reactions
 and beta-processes (including neutrino-nuclear interactions listed in Table$\,$\ref{Tab2})
 connecting 40 most important light nuclides.
 The second code (H) controls the nuclear kinetic for the heavier nuclides
 up to $Z=82$ (Pb). In our calculations the H code deals with about 1300 isotopes.
 It uses effective method of Gear \cite{Gear71} for solving stiff
 systems of differential nuclear-kinetic equations and involves a special algorithm
 of converting the sparse Jacobian matrix (see detailed description
 in \cite{BlnPan96}).

 Both the codes work consistently by iterative exchange with free neutrons
 and protons. The method was first proposed in \cite{NPB98} and then was
 developed in \cite{NadPan01,NadPan01a} where one can find further details.

\section*{Results and discussion}
We have fulfilled the calculations for different values
of the mixing parameter $\alpha=$ 0, 0.3, 0.4, 0.5, and 1.

\begin{figure}[htb!]
\centerline{\includegraphics[clip,width=0.5\textheight]{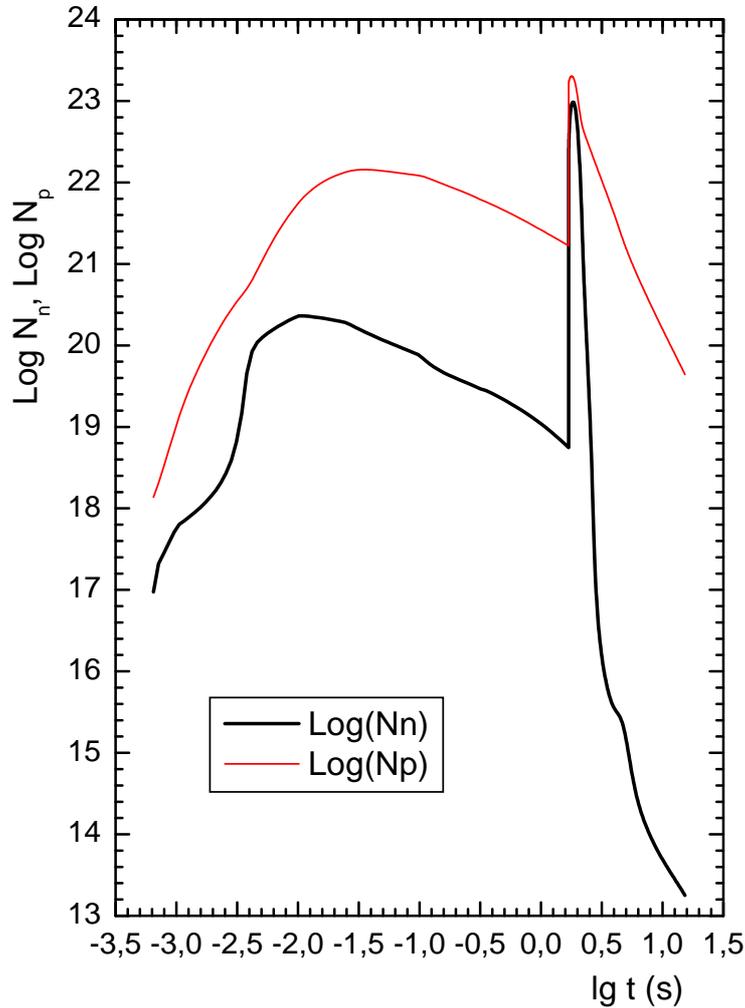}}
  \caption{The free neutron N$_\mathrm{n}$ and proton N$_\mathrm{p}$
  densities (in cm$^{-3}$) versus the
  time (in logarithmic scale) measured from the beginning of the collapse
  at $t=0$.
  }
  \label{NnNp}
\end{figure}
Figure$\,$\ref{NnNp} shows the densities of free neutrons N$_\mathrm{n}$
and protons N$_\mathrm{p}$ as functions of time for the case $\alpha=1$.
Before the SW arrival ($0<t< 1.688\,$s), the species
n, p, \isn{He}{3}, and \isn{H}{3} produced in the first two reactions
of Table$\,$\ref{Tab2} interact with
themselves, \isn{He}{4}, admixtures of the C, N, O, Ne isotopes listed in
Table$\,$\ref{Tab1}, and with ``Fe'' peak nuclides.

As a result, considerable part of neutrons released in the reaction
\isn{He}{4}($\nu,\nu^\prime$ n)\isn{He}{3} is absorbed in the reaction
\isn{He}{3}(n,p)\isn{H}{3} followed by a chain of thermonuclear reactions
creating light nuclides with excess of neutrons, such as \isn{C}{13,14},
\isn{N}{15}, \isn{O}{17,18}, \isn{F}{19}, \isn{Ne}{21} in amounts
comparable with that of Fe-seeds. What why N$_\mathrm{n}(t)$
begins to decrease after attaining its maximum at $t\approx 0.01\,$s.

However, at high temperature after the SW arrival, above nuclides
undergo a quick destruction
through the reactions \isn{C}{13}(\isn{He}{4}, n)\isn{O}{16},
\isn{C}{14}(p, n)\isn{N}{14}, \isn{O}{17}(\isn{He}{4}, n)\isn{Ne}{20},
\isn{O}{18}(\isn{He}{4}, n)\isn{Ne}{21}, \isn{F}{19}(p, n)\isn{Ne}{19},\\
\isn{Ne}{21}(\isn{He}{4}, n)\isn{Mg}{24} which emit neutrons
forming a large maximum of N$_\mathrm{n}(t)$ in Fig$\,$\ref{NnNp}.

\begin{figure}[htb]
  \centerline{\includegraphics[clip,width=0.8\textwidth]{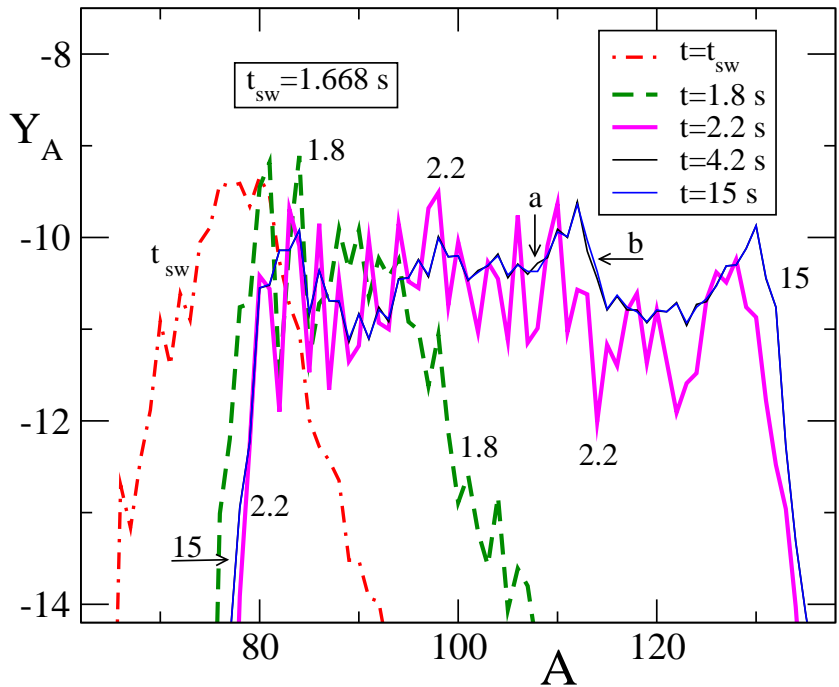}}
  \caption{The temporal development of the r-process for $\alpha =1$.}
  \label{rpt_alpha1}
\end{figure}

Figure$\,$\ref{rpt_alpha1} shows
the r-process abundances of heavy nuclides
$Y_A = \nucmu N_A/\rho$ versus mass number $A$ for several times.
Since the neutron and proton captures do not change
the total number of heavy nuclides,
at any time  the $Y_A$  distribution  meets an equation
$\sum_A Y_A = Y_{\mathrm{Fe}}$ where
$Y_{\mathrm{Fe}}= \nucmu N_{\mathrm{Fe}}/
\rho = X_{\mathrm{Fe}}/56 = $\xmn{2.46}{-9} is the initial (at t=0) value.

The $Y_A$ distribution for $t=4.2\,$s is virtually indistinguishable from
that for $t=15\,$s. One can observe only tiny differences marked by arrows
with letters ``a'' and ``b''.
This means that the r-process comes to its end at
$t\sim 4\,$s when temperature and density fall below \xmn{5.7}{8}K
and \xmn{8.2}{3}g$\,$cm$^{-3}$, respectively.
The radius of the expanding shell increases by a factor of 3 and
the neutrino flux decreases by an order of magnitude.
In total only \xmn{1.6}{53}$\,$erg or $\sim 30$\% of available neutrino
energy was actually used in our calculations.

\begin{figure}[htb!]
  \centerline{\includegraphics[clip,width=0.8\textwidth]{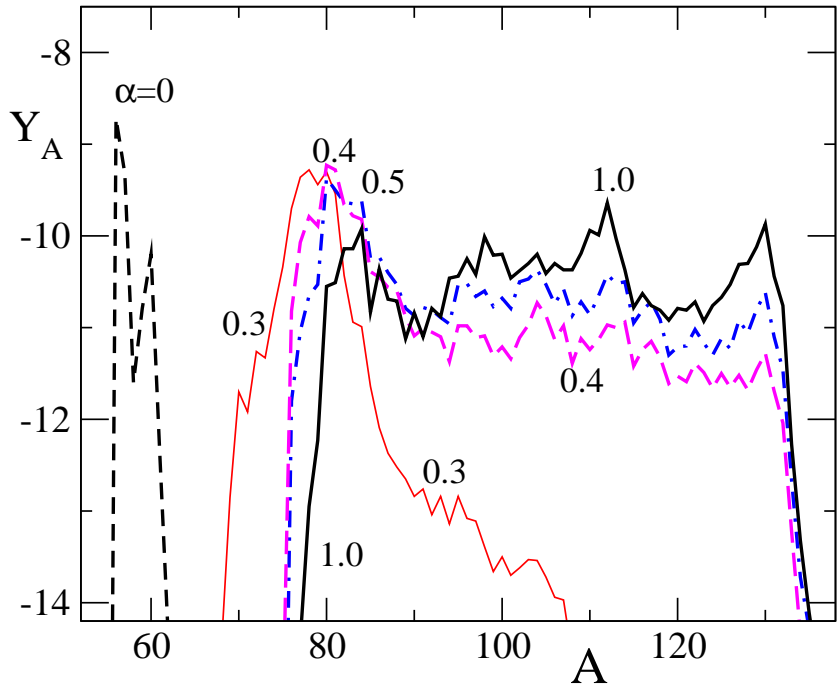}}
  \caption{The final yields at $t=15\,$s for different degree of mixing $\alpha$.}
  \label{finyields}
\end{figure}

The final yields are shown in Fig.$\,$\ref{finyields}
for different degree of mixing $\alpha$.
One can observe that $\alpha$ should exceed a critical value 0.3
for the r-process to be launched beyond $A\approx 80$.
\begin{table}[htb]
\centering
 \caption{The number of neutrons and protons captured per ``Fe'' seed}\label{Tab3}
\vspace{0.2cm}
\begin{tabular}{lllllc}
\hline\hline\\[-0.32cm]
 \multicolumn{3}{c}{\quad Before SW $(t=t_\mathrm{SW}$)}
 & \multicolumn{3}{c}{Final ($t=15\,$s)} \\[1mm]
 $\alpha$ & \multicolumn{1}{c}{n/``Fe''} & \multicolumn{1}{c}{p/``Fe''}
 & \multicolumn{1}{c}{n/``Fe''} & \multicolumn{1}{c}{p/``Fe''}
 & \multicolumn{1}{c}{(n+p)/``Fe''}\\ \hline
     & & & & &\\[-0.32cm]
 0.0 & 0.240 &\xmn{1.7}{-8} & 0.234 & 0.067 & 0.301\\
 0.3 & 6.0   &\xmn{6.7}{-5} & 20.4  & 0.7 & 21.1\\
 0.4 & 7.0   &\xmn{1.1}{-4} & 31.0  & 1.8 & 32.8\\
 0.5 & 11.8  &\xmn{2.0}{-4} & 32.9  & 2.72& 35.6\\
 1.0 & 21.0  &\xmn{4.4}{-4} & 46.8  & 2.73& 49.5\\ \hline
           & & & &   \\[-3mm]
 1.0$^a$ & 21.0  & 0.0          & 27.3  & 0.0 & 27.3\\
 1.0$^b$ & 21.8  & \xmn{5.6}{-4}  & 62.3  & 3.5 & 65.8\\[-5mm]
          & & & & \\ \hline\hline
\end{tabular}
\end{table}

Table$\,$\ref{Tab3} sum up the results of our calculations for
different $\alpha$ in terms of the numbers of neutrons and protons
captured by all heavy nuclides per ``Fe'' seed. The second and third
columns give the results at the moment of the shock arrival whereas
the final values are in next two columns. The last column shows
the total number of nucleons (n+p) captured per ``Fe'' seed.

Note that in case $\alpha =0$ the final value of n/``Fe'' at
$t=15\,$s is a little larger than that at $t=t_\mathrm{SW}$.
This occurs due to the (p,$\,$n) reactions activated on heavy nuclides
after the SW arrival.

The last line of Table$\,$\ref{Tab3} is for the case
when the neutrino luminosity curve is described by the law
$L_\nu\sim\exp(-t/\tau)$ with the e-folding time $\tau =3\,$s used,
for example, in \cite{BanHaxQuian11}. Such a fast decay
of $L_\nu$ contradicts to the observations of supernova 1987A
in the Large Magellanic Cloud. However, it increases the fraction
of the total
radiated neutrino energy that could be useful for the nucleosynthesis
we deal with. At obtained above quenching time 4.2$\,$s about
75\% of all neutrino energy is radiated rather than only 30\% for
the neutrino light curve adopted in our calculations.

\begin{figure}[htb!]
  \centerline{\includegraphics[clip,width=0.8\textwidth]{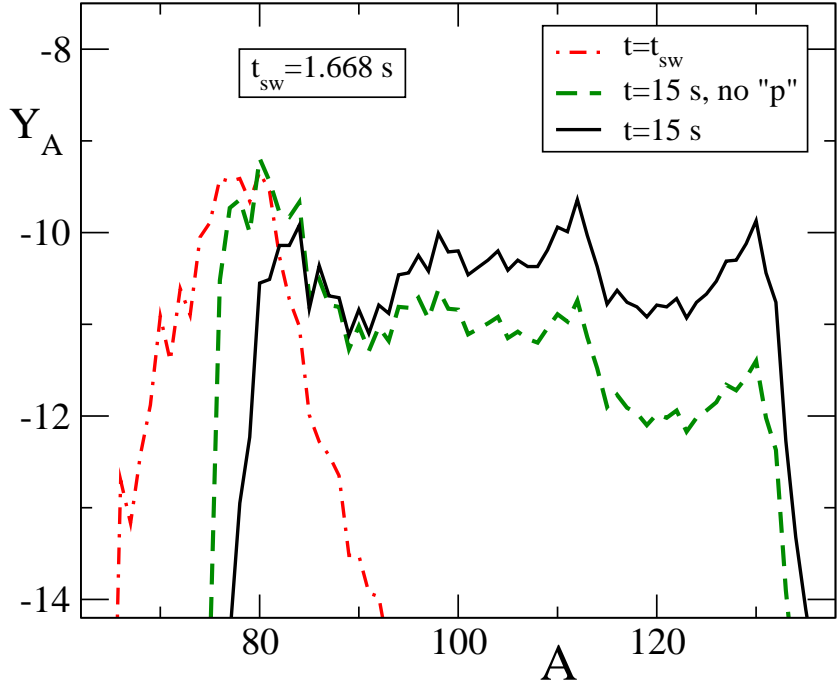}}
  \caption{The contribution of the proton capture reactions to the r-process.}
  \label{p_efficiency}
\end{figure}

Figure$\,$\ref{p_efficiency} demonstrates the role of the proton
capture reactions (p,$\gamma$) and (p,n) by heavy nuclei
created from ``Fe'' seeds.
These reactions, activated by high temperature of shocked matter,
increase atomic number $Z$ and thereby accelerate the r-process.
The dash-dotted and solid lines are from Fig.$\,$\ref{rpt_alpha1}.
The dashed line comes from the calculation neglecting above proton
capture reactions.
A comparison of the solid and dashed lines shows that the abundance
peaks at $A\approx\,$110 and 130 become about an order of magnitude higher
due to the proton capture reactions. This result is in accord
with conclusion of our previous work \cite{PanNad99}.
The total number
of nucleons captured per ``Fe'' seed turns out to be by a factor
of 1.8(!) lower than in case when the proton capture reactions
are taken into account (the last but one line in Table$\,$\ref{Tab3}).

\section*{Conclusion}
    To guarantee a noticeable contribution of the neutrino radiation
    into the weak r-process component the following conditions
    have to be satisfied.

1. The breaking of spherical symmetry is
    necessary in presupernova models to take  into
    account possible large-scale circulating currents
    transporting some He from $R=\xmn{4}{9}$-- \xmn{4}{10}cm
    in deeper layers.

2. Helium  has to be at radius $R\lesssim 10^9$cm
   with abundance by weight $X_{\mathrm He}\gtrsim 0.3$.

3. Shock wave must be described accurately
    since it activates evaporation of neutrons
    absorbed by light elements before its arrival  as well as stimulates
    the proton (p,n), (p,$\gamma$) captures by heavy elements produced
    from ``Fe'' seeds, thereby considerably speeding up the r-process.

\vspace*{3mm}
 {\bf Acknowledgement}.

  Authors are grateful to A.~Heger for supplying us with presupernova
  model dd15z-4.

  The work was finished in Physical Department
  of the University of Basel. We thank Prof. F.--K. Thielemann
  for hospitality and support.

  The work was supported
  by the Russian Foundation Basic Research (RFBR) grants
  11-02-00882-a, 12-02-00955-а, the Russian government
  grant 11.G34.31.0047 and the Swiss National Science Foundation
  SCOPES project No.~IZ73Z0-128180/1.

\end{document}